\documentclass{ws-ijmpcs}

\def\bce{\begin{center}}
\def\ece{\end{center}}
\def\beq{\begin{eqnarray}}
\def\eeq{\end{eqnarray}}
\def\ben{\begin{enumerate}}
\def\een{\end{enumerate}}
\def\bei{\begin{itemize}}
\def\eei{\end{itemize}}

\def\nn{\nonumber}

\def\tr{\mbox{tr\,}}
\def\Tr{\mbox{Tr\,}}
\def\Re{\mbox{Re\,}}

\def\brr{\begin{array}}
\def\err{\end{array}}
\def\dsp{\displaystyle}

\begin{document}

\markboth{Emilio Elizalde}
{Zeta function regularization in Casimir effect calculations and J.S. Dowker's contribution}

\catchline{}{}{}{}{}
%
%%%%%%%%%%%%%%%%%%%%%%%%%%%%%%%%%%%%%%%%%%%%%%%%%%%%%%%%%%%%%%%%%%%%

\title{ZETA FUNCTION REGULARIZATION
IN CASIMIR EFFECT CALCULATIONS
AND J. S. DOWKER's CONTRIBUTION}

\author{EMILIO ELIZALDE}

\address{Consejo Superior de Investigaciones Cient\'\i ficas,
ICE-CSIC and IEEC \\ Campus UAB, Facultat de Ci\`encies, Torre
C5-Parell-2a pl \\ 08193 Bellaterra (Barcelona) Spain\\
 elizalde@ieec.uab.es, elizalde@math.mit.edu\\
www.ieec.cat/english/recerca/ftc/eli/eli.htm}

\maketitle

\begin{history}
\received{Day Month Year}
\revised{Day Month Year}
\end{history}

\begin{abstract}
A summary of relevant contributions, ordered in time, to the subject of operator zeta functions
and their application to physical issues is provided. The description ends with the seminal contributions of Stephen Hawking and Stuart Dowker and collaborators, considered by many authors as the actual starting point of the introduction of zeta function regularization methods in theoretical physics, in particular, for quantum vacuum fluctuation and Casimir effect calculations.
After recalling a number of the strengths of this powerful and elegant method, some of its limitations are discussed. Finally, recent results of the so called operator regularization procedure are presented.

\end{abstract}

\ccode{PACS numbers: 98.80.Jk, 11.10.Gh}

\section{Introduction}\label{aba:sec1}
We devote this introductory section to a historical account of some relevant contributions to the subject of operator zeta functions and their applications in theoretical physics. We start recalling the definition of the standard zeta function:
the Riemann zeta function, $\zeta (s)$, which is a function of a complex variable, $s$. To define it, one considers the infinite series
\beq  \sum_{n=1}^\infty \frac{1}{n^s}\eeq
 which is absolutely convergent for all complex values of $s$ such that Re $ s> 1$, and then defines $\zeta (s)$ as the analytic continuation, to the whole complex $ s-$plane, of the function given for Re $ s> 1$ by the sum of the preceding series. Actually, Leonhard Euler had already considered the above series in 1740, but only for positive integer values of $ s$ and some years later Chebyshev had extended the definition to Re $  s > 1$. Riemann formulated his famous hypothesis on the non-trivial zeros of the zeta function in 1859\cite{rie1}. It turns out that this is the only one in the famous list of twenty-three problems discussed in the address given in Paris by David Hilbert (on August 9, 1900), which has gone into the new list of seven Millennium Prize Problems, established by the Clay Mathematics Institute of Cambridge, Massachusetts (USA), and which were announced at a meeting in Paris (held on May 24, 2000) at the Coll\`{e}ge de France.

In 1916, in their seminal paper {\it ``Contributions to the Theory of the Riemann Zeta-Function and the Theory of the Distribution of Primes"}\cite{hl1}, Godfrey H. Hardy and John E. Littlewood
did much of the earlier work concerning possible applications of the zeta function as a regularization procedure, by establishing the convergence and equivalence of series regularized with the heat kernel and zeta function regularization methods. Also very important in this respect was the appearance of Hardy's book entitled {\it Divergent Series}\cite{har1}. We should also note that Srinivasa I. Ramanujan had already found, working in isolation, the functional equation of the zeta function, independently of all this development, as Hardy could later certify.

Torsten Carleman, in 1935, in his work in French {\it Propri\'{e}t\'{e}s asymptotiques des fonctions fondamentales des membranes vibrantes}\cite{carl1}, obtained the zeta function encoding the eigenvalues of the Laplacian of a compact Riemannian manifold for the case of a compact region of the plane. This was an important first step towards extending the concept of zeta function as associated to the spectrum of a differential operator, which is actually the situation at issue here. And, as has been much more widely recognized in the literature, a decade and a half later Subbaramiah Minakshisundaram and {\AA}ke Pleijel, in their 1949 paper  {\it Some properties of the eigenfunctions of the Laplace-operator on Riemannian manifolds}\cite{mp1}, extended Carleman's results explicitly showing that, if $ A$ is the Laplacian of a compact Riemannian manifold, then the corresponding zeta function, $\zeta_A(s)$, converges and has an analytic continuation as a meromorphic function to all complex numbers, what is actually a very remarkable result.

Another milestone in this development was Robert Seeley's seminal work, published in 1967, {\it Complex powers of an elliptic operator}\cite{seel1}. Seeley fully extended in this paper the above treatment to general {\it elliptic pseudo-differential} operators on compact Riemannian manifolds. He proved that, for all such operators, one can rigorously define a determinant using zeta function regularization. In 1971, Daniel B. Ray and Isadore M. Singer\cite{rays1} used Seeley's theory in their famous paper entitled {\it $R$-torsion and the Laplacian on Riemannian manifolds} to define the determinant of a positive self-adjoint operator, $A$. Such operator is, in their explicit applications, the Laplacian of a Riemannian manifold; denoting its eigenvalues by $a_1, a_2, ....,$ then its zeta function is formally given by the trace
\beq \zeta_A (s) =  \Tr (A^{-s}). \eeq
The method defines also the (possibly divergent) infinite product as
\beq \prod_{n=1}^\infty a_n =  \exp [-{\zeta_A}'(0)]. \eeq

At this point we arrive, in our chronological survey, to the very important contribution of Stuart Dowker and Raymond Critchley. In their seminal work, published in 1976, {\it Effective Lagrangian and energy-momentum tensor in de Sitter space}\cite{dowc1}, these authors went definitely further in the application of the above procedures to physics: they actually proposed, for the first time, a fully-fledged zeta function regularization method for quantum physical systems. This paper has got high recognition, having gathered over 600 citations to present date. For the sake of completeness let us here reproduce in full its abstract:

{\it The effective Lagrangian and vacuum energy-momentum tensor $< T^{\mu\nu}>$ due to a scalar field in a de Sitter space background are calculated using the dimensional-regularization method. For generality the scalar field equation is chosen in the form  $(\Box^2+ \xi R + m ^2)\varphi = 0$. If $\xi = 1/6$ and $m = 0,$ the renormalized $<T^{\mu\nu}>$ equals $g^{\mu\nu}(960\pi^2a^4)^{-1},$ where $a$ is the radius of de Sitter space. More formally, a general zeta-function method is developed. It yields the renormalized effective Lagrangian as the derivative of the zeta function on the curved
space. This method is shown to be virtually identical to a method of dimensional regularization applicable to any Riemann space.}

One thing specialists often point out is that, in spite of the fact that, elaborating from the methods developed in this paper, it is true that a well defined and clear regularization prescription for a general case can be easily obtained, the authors actually described the method very briefly in this work, the uses and wide possibilities of the procedure not having been fully exploited or even anticipated there.

This is maybe the main reason why Stephen Hawking's 1977 extremely influential paper (it has got over 1100 citations up to date) entitled {\it Zeta function regularization of path integrals in curved spacetime}\cite{hawk1} is considered by many to be the actual seminal reference where the zeta function regularization method was defined, with all its computational power and possible physical applications, which were very clearly identified there. Needless to say, the title of the paper is absolutely explicit. Again, let us reproduce, for comparison, its abstract:

{\it This paper describes a technique for regularizing quadratic path integrals on a curved background spacetime. One forms a generalized zeta function from the eigenvalues of the differential operator that appears in the action integral. The  zeta function is a meromorphic function and its gradient at the origin is defined to be the determinant of the operator. This technique agrees with dimensional regularization where one generalises to $  n$ dimensions by adding extra flat dimensions. The generalized zeta function can be expressed as a Mellin transform of the kernel of the heat equation which describes diffusion over the four dimensional spacetime manifold in a fifth dimension of parameter time. Using the asymptotic expansion for the heat kernel, one can deduce the behaviour of the path integral under scale transformations of the background metric. This suggests that there may be a natural cut off in the integral over all black hole background metrics. By functionally differentiating the path integral one obtains an energy momentum tensor which is finite even on the horizon of a black hole. This electromagnetic tensor has an anomalous trace.}

In my view, after investigating the case in some detail, it is fair to conclude that the priority of Dowker and Critchley in this matter has been sufficiently well established in the literature I have consulted (with some really incredible exceptions, however, as the running Wikipedia article on {\it ``Zeta function regularization"}, where not the least reference to Dowker and Critchley is done!). Further to this, considering the number of citations collected by each one of the two papers, taking then into account the enormous mediatic impact of S.W. Hawking to modern physics (and well beyond it), and also the careful analysis of both the abstracts and the whole papers themselves, the ratio of citations to both works seems fair enough. But this is just to be taken as my personal opinion, of course.

To continue this account further would require a very hard work and would end in a very long report, at least book size, what is not the purpose here. Let us finish this short report here, at the point when, as already mentioned, the zeta function regularization method is considered to have been clearly defined and its usefulness for physics undoubtedly established. A large number of very interesting research articles in several directions have been published on these matters during the last 35 years. I will just mention a few references [\refcite{zbooks1}], which by no means are meant to constitute an optimized list. In the next section a short description of the main basic concepts involved in any rigorous formulation of the procedure of zeta function regularization will be given, together with some results originally obtained by the author.

\section{Zeta function of a pseudodifferential operator and determinant}

\subsection{The zeta function}
The {\it zeta function} $\zeta_A$ of $A$, a positive-definite elliptic pseudodifferential operator ($\Psi$DO) of positive order $m \in \mbox{\bf R}$ (acting on the space of smooth sections of
$E$, an $n$-dimensional vector bundle over
 a  closed $n$-dimensional manifold, $M$)  is defined as
\beq
\zeta_A (s) = \mbox{tr}\  A^{-s} = \sum_j
 \lambda_j^{-s}, \qquad \mbox{Re}\ s>\frac{n}{m} \equiv s_0.
\eeq
being
$s_0=$ dim$\,M/$ord$\,A$ the {\it abscissa of
convergence} of $\zeta_A(s)$.
It can be proven that $\zeta_A(s)$ has a meromorphic
continuation to  the whole complex plane
$\mbox{\bf C}$ (regular at $s=0$),
provided the principal symbol of $A$ ($a_m(x,\xi)$) admits a {\it spectral cut}: $
L_\theta = \left\{  \lambda \in \mbox{\bf C};
 \mbox{Arg}\, \lambda =\theta,
\theta_1 < \theta < \theta_2\right\}$,   $\mbox{Spec}\, A
\cap L_\theta = \emptyset$
(Agmon-Nirenberg condition\cite{kont95b}).
 This definition of $\zeta_A (s)$ depends on the
position of the cut $L_\theta$.
 The {\it only possible} singularities of $\zeta_A (s)$ are
{\it simple poles} at
$
s_k = (n-k)/m,  \  k=0,1,2,\ldots,n-1,n+1, \dots.
$
 M. Kontsevich and S. Vishik have managed to extend this
definition to the case when  $m \in \mbox{\bf C}$ (no spectral
cut exists)\cite{kont95b}.

\subsection{Zeta regularized determinant}

Let $A$ be a $\Psi$DO  operator with a  spectral decomposition:
$ \{ \varphi_i, \lambda_i \}_{i\in I}$, with $I$ some set of indices.
The definition of determinant starts by trying
to make sense of the product $\prod_{i\in I} \lambda_i$, which can be
easily transformed into a ``sum'':
   $ \ln \prod_{i\in I} \lambda_i   = \
      \sum_{i\in I} \ln \lambda_i $. From  the
definition of the  zeta function of $A$:  $\zeta_A (s) =
    \sum_{i\in I} \lambda_i^{-s}$, by
 taking the derivative at $s=0$:   $\zeta_A ' (0) =  - \sum_{i\in I}
   \ln \lambda_i $, we arrive to the following definition of determinant
of $A$ [\refcite{rs1}]:
  \beq {\det}_\zeta A = \exp \left[ -\zeta_A ' (0)
   \right].
   \eeq
An older definition (due to Weierstrass) is obtained by subtracting
in the series above (when it is such) the leading behavior of
 $\lambda_i$ as a function of  $i$, as $i \rightarrow \infty$,
 until the series
 $ \sum_{i\in I} \ln \lambda_i $ is made to converge [\refcite{bs01}]. The shortcoming
---for physical applications--- is here that these additional terms turn out to be
{\it non-local} and, thus, are non-admissible in a renormalization procedure.

In algebraic QFT, to write down an action in operator language
one needs a functional that replaces integration.
 For the Yang-Mills theory this is the Dixmier trace, which
 is the {\it unique} extension of the usual trace to the ideal
 ${\cal L}^{(1,\infty)} $
of the compact operators  $T $ such that the partial
 sums of its spectrum diverge
logarithmically as the number of terms in the sum:
$
\sigma_N (T) \equiv \sum_{j=0}^{N-1} \mu_j= {\cal O}
(\log N), \
\mu_0 \geq \mu_1 \geq \cdots$
 The definition of the Dixmier trace of $T$ is: Dtr $T =
 \lim_{N\rightarrow \infty}
\frac{1}{\log N} \sigma_N (T),$
provided that the Cesaro means $M(\sigma) (N)$
 of the sequence in $N$ are convergent as $N
 \rightarrow \infty$
(remember that: $M(f)(\lambda) =\frac{1}{\ln \lambda}
\int_1^\lambda f(u) \frac{du}{u}$).
 Then, the Hardy-Littlewood theorem can be stated in a way
 that connects the Dixmier trace with the residue of the zeta
function of the operator $T^{-1}$ at $s=1$ (see Connes [\refcite{conn1}]):
$ \mbox{Dtr}\ T=
\lim_{ s \rightarrow  1^+} (s-1) \zeta_{T^{-1}} (s). $

The Wodzicki (or noncommutative) residue\cite{wodz87b}
 is the {\it only} extension of the Dixmier trace to the
$\Psi$DOs which are not
in  ${\cal L}^{(1,\infty)}$.
It is the {\it only}
 trace  one can define in the algebra of $\Psi$DOs (up to a
multiplicative constant),
 its definition being: res $A=2$ Res$_{s=0}\ \tr (A \Delta^{-s})$,
with  $\Delta$ the Laplacian. It
satisfies the trace condition: res $(AB)=$ res $(BA)$. A very
important property is that  it can be expressed as an integral (local form)
$
\mbox{res} \ A = \int_{S^*M} \mbox{tr}\   a_{-n}(x,\xi)
 \, d\xi
$
with $S^*M \subset T^*M $ the co-sphere bundle on $M$
(some authors put a coefficient  in front of the integral: Adler-Manin
residue\cite{adman1}).

If dim $M=n=-$  ord $A$ ($M$ compact Riemann,
$A$ elliptic, $n\in \mbox{\bf N}$) it coincides with
 the Dixmier trace, and one has
$
\mbox{Res}_{s=1} \zeta_A (s) = \frac{1}{n} \, \mbox{res} \ A^{-1}.
$
 The Wodzicki residue continues to make sense
for $\Psi$DOs of arbitrary order and, even if the symbols
$a_{j} (x, \xi)$, $j<m$, are not invariant under
coordinate choice,
their integral is, and defines a trace. All residua at poles
of the zeta function of a $\Psi$DO can be easily obtained
from the Wodzciki residue\cite{6a}.
%%%%%%%%%%%%%%%%%%%%%%%%%%%%%%%%
\subsection{Multiplicative anomaly}
%%%%%%%%%%%%%%%%%%%%%%%%%%%%%%%%
Given $A$, $B$ and $AB$ $\Psi$DOs, even if
$\zeta_A$, $\zeta_B$ and $\zeta_{AB}$ exist, it turns out that, in
general,
$
{\det}_\zeta (AB) \neq {\det}_\zeta A \ {\det}_\zeta B.
$
 The multiplicative (or noncommutative, or determinant) anomaly is defined as:
\beq
\delta (A,B) = \ln \left[ \frac{
\det_\zeta (AB)}{\det_\zeta A \ \det_\zeta B} \right] =
-\zeta_{AB}'(0)+\zeta_A'(0)+\zeta_B'(0).
\eeq
 Wodzicki's formula for the multiplicative anomaly\cite{wodz87b,kass1}:
\beq
\delta (A,B)= \frac{\mbox{res}\left\{ \left[ \ln \sigma (A,B)\right]^2
\right\}}{2 \mbox{ord}\, A  \mbox{ord}\, B
(\mbox{ord}\, A + \mbox{ord}\, B)}, \quad
\sigma (A,B) := A^{\mbox{ord}\, B} B^{-\mbox{ord}\, A}.
\eeq

At the level of Quantum Mechanics (QM), where it was originally introduced by
Feynman, the path-integral approach is just an
alternative formulation of the theory. In QFT it is
much more than this, being in many occasions {\it the} actual formulation
of QFT [\refcite{ramond}]. In short, consider the Gaussian functional integration
\beq
\int [d\Phi ] \ \exp \left\{ - \int d^D x \left[ \Phi^\dagger (x) \big(
\ \ \ \ \big) \Phi (x) + \cdots \right] \right\}  \quad
  \longrightarrow  \quad
\det \big( \  \ \big)^{\pm 1}, \eeq
(the sign $\pm$ depends on the spin-class of the integration fields) and assume that the operator matrix has the following simple structure
(being each $A_i$ an operator on its own):
\beq  \left( \brr{cc} A_1 & A_2 \\ A_3 & A_4 \err \right)
\qquad \longrightarrow  \quad
 \left( \brr{cc} A & \mbox{} \\ \mbox{} & B \err \right), \eeq
 where the last expression is the result of diagonalizing the
operator matrix.
A question now arises. What is the determinant of the operator matrix:
$ \det (AB)$ or $\det A  \cdot \det B$?
This has been very much on discussion during the last months\cite{1}.
There is agreement in that:
(i)  In a situation where a superselection rule exists, $AB$
has no sense (much less its determinant), and then the answer must be $
\det A \cdot \det B$.
(ii) If the diagonal form is obtained after a change of basis
(diagonalization process), then the quantity that is preserved by
such transformations is the value of  $
\det (AB)$ and {\it not} the product of the individual determinants
(there are counterexamples supporting this viewpoint\cite{2}).
%%%%%%%%%%%%%%%%%%%%%%%%%%%%%%%%
\section{Explicit calculation of $\zeta_A$ and $\det_\zeta A$}
%%%%%%%%%%%%%%%%%%%%%%%%%%%%%%%%
A fundamental property of many zeta functions is the existence
of a reflection formula, also known as the functional equation by 
mathematicians. For the   Riemann zeta function:
  $\Gamma (s/2) \zeta (s)
=\pi^{s-1/2} \Gamma (1-s/2) \zeta (1-s)$.
For a generic zeta function, $Z(s)$, it is
$
Z(\omega -s)= F(\omega, s) Z(s),
$
 and allows for its analytic continuation in an
easy way ---what is, as advanced above,
 the whole story of the zeta function regularization
 procedure (at least the main part of it). But the
analytically continued expression thus obtained
 is just another series,  again with
a slow convergence behavior, of power series type\cite{bo1} (actually the same
that the original series had,
in its own domain of validity). S. Chowla and A. Selberg
found a formula, for the Epstein zeta function in the two-dimensional case\cite{cs},
that yields {\it exponentially quick convergence, and not only in the reflected
domain}. They were extremely proud of that formula ---as one can
appreciate just reading the original paper (where actually no hint about its
 derivation was given, see [\refcite{cs}]).
In Ref.~[\refcite{eli2a}], I generalized this expression to inhomogeneous
zeta functions (most important for physical applications),
but staying always in {\it two} dimensions,
for this was commonly believed to be an
unsurmountable restriction of the original formula (see, e.g.,
Ref.~[\refcite{dic}]). Later I obtained an extension to an {\it arbitrary} number
of dimensions\cite{cmpe1}, both in the homogeneous (quadratic form) and non-homogeneous
(quadratic plus affine form) cases.

In short, for the following zeta functions (corresponding
to the general quadratic ---plus affine--- case and to the general affine
case, in any number of dimensions, $d$) explicit formulas of the CS type
were obtained in [\refcite{cmpe1}], namely,
\beq \zeta_1 (s) = \sum_{\vec{n} \in \mbox{\bf Z}^d} [Q(\vec{n})
+A(\vec{n})]^{-s}\eeq and \beq  \zeta_2 (s) = \sum_{\vec{n} \in \mbox{\bf N}^d}
A(\vec{n})^{-s},\eeq  where $Q$ is a non-negative quadratic form and $A$ a general
affine one, in $d$ dimensions (giving rise to Epstein and Barnes zeta
functions, respectively). Moreover, expressions for the more
difficult cases when the summation ranges are interchanged, that is:
\beq  \zeta_3 (s) = \sum_{\vec{n} \in \mbox{\bf N}^d} [Q(\vec{n})
+A(\vec{n})]^{-s}\eeq  and \beq  \zeta_4 (s) = \sum_{\vec{n} \in \mbox{\bf Z}^d}
A(\vec{n})^{-s}\eeq  have been given in [\refcite{cmpe1}].
%%%%%%%%%%%%%%%%%%%%%%%%%%%%%%%%
\subsection{Extended Epstein zeta function in $p$ dimensions}
%%%%%%%%%%%%%%%%%%%%%%%%%%%%%%%%
The starting point is {\it Poisson's resummation formula} in $p$
 dimensions, which arises from the distribution identity
$
\sum_{\vec{n} \in \mbox{\bf Z}^p} \delta (\vec{x}-\vec{n}) =
\sum_{\vec{m} \in \mbox{\bf Z}^p} e^{i 2\pi \vec{m} \cdot \vec{x}}.
\label{prs1}
$
(We shall indistinctly write $ \vec{m} \cdot \vec{x} \equiv
 \vec{m}^T \vec{x}$ in what follows.) Applying this identity to the
function
$
f(\vec{x}) = \exp \left( -\frac{1}{2}\vec{x}^T A  \vec{x} +
 \vec{b}^T \vec{x} \right),
$
with $A$ an invertible $p\times p$ matrix, and integrating  over
$\vec{x} \in \mbox{\bf R}^p$, we get:
$
\sum_{\vec{n} \in \mbox{\bf Z}^p} \exp
\left( -\frac{1}{2}\vec{n}^T A  \vec{n} +
 \vec{b}^T \vec{n} \right) = \frac{(2 \pi)^{p/2}}{\sqrt{\det A}}
\sum_{\vec{m} \in \mbox{\bf Z}^p}
 \exp \left[ \frac{1}{2}\left(\vec{b} + 2\pi i \vec{m} \right)^T A^{-1}
\left(\vec{b} + 2\pi i \vec{m} \right)\right]. \nn %\label{prs2}
$
Consider now the following zeta function ($\Re s > p/2$):
\beq
\hspace*{-3mm} \zeta_{A,\vec{c},q} (s) = {\sum_{\vec{n} \in \mbox{\bf Z}^p}}'
 \left[
\frac{1}{2}\left( \vec{n}+\vec{c}\right)^T A
\left( \vec{n}+\vec{c}\right)+ q\right]^{-s} \equiv
{ \sum_{\vec{n} \in \mbox{\bf Z}^p}}'
\left[ Q\left( \vec{n}+\vec{c}\right)+ q\right]^{-s} \nn % \label{zf1}
\eeq
(the prime on the summation signs mean
that the point $ \vec{n}=\vec{0}$ is to be excluded from the sum).
The aim is to obtain a formula giving (the analytic continuation
of) this multidimensional
zeta function in terms of an exponentially convergent multiseries
and which is valid in
the whole complex plane, explicitly exhibiting the singularities (simple poles) of
the meromorphic continuation and the corresponding residua.
 The only condition on the matrix $A$ is that
it corresponds to a (non negative) quadratic form, which we call $Q$.
The vector $\vec{c}$ is arbitrary, while $q$ will (for the
moment) be a positive constant.
Use of the Poisson resummation formula yields
\beq
\zeta_{A,\vec{c},q} (s) &=& \frac{(2\pi )^{p/2} q^{p/2 -s}}{
\sqrt{\det A}} \, \frac{\Gamma(s-p/2)}{\Gamma (s)} +
 \frac{2^{s/2+p/4+2}\pi^s q^{-s/2 +p/4}}{\sqrt{\det A} \
\Gamma (s)}  \label{qpd1}\\ && \hspace*{-20mm} \times {\sum_{\vec{m} \in
\mbox{\bf Z}^p_{1/2}}}' \cos
(2\pi
 \vec{m}\cdot \vec{c}) \left( \vec{m}^T A^{-1} \vec{m}
\right)^{s/2-p/4} \, K_{p/2-s} \left( 2\pi \sqrt{2q \,
 \vec{m}^T A^{-1} \vec{m}}\right), \nn
\eeq
where $K_\nu$ is the modified Bessel function of the second kind, and
the subindex 1/2 in
$\mbox{\bf Z}^p_{1/2}$ means that only half of the vectors
$\vec{m} \in \mbox{\bf Z}^p$ intervene in the sum. That is, if we take
an $\vec{m} \in \mbox{\bf Z}^p$
we must then exclude $-\vec{m}$ (as simple criterion one can,
for instance, select those vectors in {\bf Z}$^p \backslash \{ \vec{0}
\}$ whose first non-zero component is positive). Eq. (\ref{qpd1})
fulfills {\it all} the requirements demanded before. It
is notorious to observe how the only pole of this inhomogeneous Epstein
zeta function appears explicitly at $s=p/2$, just where it should, its
residue being
$
\mbox{Res}_{s=p/2}  \zeta_{A,\vec{c},q} (s) =
\frac{(2\pi )^{p/2}}{\sqrt{\det A} \ \Gamma(p/2)}.
$
It is relatively simple to obtain the limit of expression
(\ref{qpd1}) as $q\rightarrow 0$.

When $q=0$ there is {\it no} way to use the
Poisson formula
on all $p$ indices of $\vec{n}$. However, one can still use
it on {\it some} of the $p$
indices $\vec{n}$ only, say on just one of them, $n_1$.
Poisson's formula on one index
reduces to the celebrated Jacobi identity for the $\theta_3$ function,
which can be written as
$
\sum_{n=-\infty}^{\infty} e^{-(n+z)^2t} =
 \sqrt{\frac{\pi}{t}} \left[ 1+ \sum_{n=1}^{\infty}
e^{-\pi^2n^2/t } \cos (2\pi n z) \right]. %\label{tfi4}
$
 Here $z$ and
$t$ are arbitrary complex numbers,  $z,t \in$ {\bf C},  with the only
restriction that  $\Re t > 0$.
Applying this last formula to the first component, $n_1$,
 we obtain the following recurrent formula
(for the sake of simplicity
we set  $\vec{c} =\vec{0}$, but the result can
be easily  generalized to $\vec{c} \neq \vec{0}$):
 \begin{eqnarray}
&& \hspace{-8mm}  \zeta_{A,\vec{0},q} (s) = \zeta_{a,\vec{0},q} (s) +
 \sqrt{\frac{\pi}{a}}\,
\frac{\Gamma (s-1/2)}{\Gamma (s)} \,  \zeta_{\Delta_{p-1},\vec{0},q}
(s-1/2) + \frac{4 \pi^s}{a^{s/2+1/4}\, \Gamma (s)}
{\sum_{\vec{n}_2 \in \mbox{\bf Z}^{p-1}}}'  \label{qrpd}  \\
&&  \hspace{-8mm}
\sum_{n_1=1}^\infty
 \cos \left( \frac{\pi n_1}{a}  \vec{b}^T  \vec{n}_2 \right)
n_1^{s-1/2}  \left( \vec{n}_2^T \Delta_{p-1}  \vec{n}_2
+q \right)^{1/4 -s/2}
 K_{s-1/2} \left( \frac{2\pi n_1}{\sqrt{a}}
\sqrt{\vec{n}_2^T \Delta_{p-1}  \vec{n}_2 +q} \right). \nn
\end{eqnarray}
This is a recurrent formula in $p$, the number of dimensions,
the first term of the recurrence being (see e.g., the 6th reference in [\refcite{zbooks1}])
\beq
\zeta_{a,\vec{0},q} (s)& =& 2 \sum_{n=1}^\infty \left(
an^2+q\right)^{-s}
= q^{-s} + \sqrt{\frac{\pi}{a}}\, \frac{\Gamma (s-1/2)}{\Gamma (s)} \,
q^{1/2 -s}\nn \\ &&
\hspace*{-10mm} + \frac{4\pi^s}{\Gamma (s)} a^{-1/4-s/2} q^{1/4-s/2}
\sum_{n=1}^\infty n^{s-1/2} K_{s-1/2} \left( 2\pi n
\sqrt{\frac{q}{a}} \right). \label{r01}
\eeq
To take in these expressions the limit $q\rightarrow 0$ is immediate:
 \begin{eqnarray}
&& \hspace*{-7mm}  \zeta_{A,\vec{0},0} (s) = 2 a^{-s} \zeta (2s) +
 \sqrt{\frac{\pi}{a}}\,
\frac{\Gamma (s-1/2)}{\Gamma (s)} \,  \zeta_{\Delta_{p-1},\vec{0},0}
(s-1/2) + \frac{4 \pi^s}{a^{s/2+1/4}\, \Gamma (s)}
{\sum_{\vec{n}_2 \in \mbox{\bf Z}^{p-1}}}' \nn \\  &&  \hspace*{-8mm}
\sum_{n_1=1}^\infty
 \cos \left( \frac{\pi n_1}{a}  \vec{b}^T  \vec{n}_2 \right)
n_1^{s-1/2}  \left( \vec{n}_2^T \Delta_{p-1}  \vec{n}_2
 \right)^{1/4 -s/2}
 K_{s-1/2} \left( \frac{2\pi n_1}{\sqrt{a}}
\sqrt{\vec{n}_2^T \Delta_{p-1}  \vec{n}_2 } \right). \label{cspd}
\end{eqnarray}
In the above formulas, $A$ is a $p \times p$ symmetric matrix
$A= \left( a_{ij} \right)_{i,j=1,2, \ldots, p} =A^T$,
$A_{p-1}$  the $(p-1) \times (p-1)$ reduced matrix
$A_{p-1}= \left( a_{ij} \right)_{i,j=2, \ldots, p}$,
$a$ the component $a=a_{11}$, $\vec{b}$ the $p-1$ vector
$\vec{b} =(a_{21}, \ldots, a_{p1})^T = (a_{12}, \ldots, a_{1p})^T$, and
 $\Delta_{p-1}$ is the following $(p-1) \times (p-1)$  matrix:
 $\Delta_{p-1} =  A_{p-1}- \frac{1}{4a} \vec{b} \otimes \vec{b}$.

It turns out  that the  limit as $q \rightarrow
0$ of Eq. (\ref{qpd1}) is again the recurrent formula
(\ref{cspd}). More precisely, what is obtained in the limit is the
reflected formula, which one gets after using Epstein zeta function's
 reflection
$
\Gamma (s) Z(s;A) = \frac{\pi^{2s-p/2}}{\sqrt{\det A}}
 \Gamma (p/2-s) Z(p/2-s;A^{-1}),
$
being $Z(s;A)$ the Epstein zeta function\cite{eps1}.
This result is easy to understand after some thinking.
 Summing up, we have thus checked that
Eq.  (\ref{qpd1}) is valid for {\it any} $q \geq 0$, since it
contains in a hidden way, for $q=0$, the recurrent expression
(\ref{cspd}).

The formulas here  can be considered as generalizations
 of the Chowla-Selberg series formula. All share the same
 properties that are so much appreciated by number-theoretists as
pertaining to the CS formula. In a way,  these expressions can be viewed
as improved reflection formulas for zeta functions; they are in fact
much better than those
in several aspects: while a reflection formula connects one
region of the complex plane with a complementary region (with some
intersection) by analytical continuation, the CS formula and the
formulas above
are  valid on the {\it whole} complex plane, exhibiting the poles
of the zeta
function and the corresponding residua {\it explicitly}. Even more important,
while a reflection formula is intended to replace the initial
expression of the zeta function ---a power series whose convergence can
be extremely slow--- by another power series with the same type of
convergence, it turns out that the expressions here obtained give the
meromorphic extension of the
zeta function, on the whole complex $s$-plane, in terms of an
{\it exponentially decreasing} power series (as was the case with the CS
formula, that one being its most precious property).
Actually, exponential convergence strictly holds
under the condition that
$q\geq 0$. However, the formulas themselves are valid for $q <0$ or even
complex. What is not guaranteed for general $q\in ${\bf C} is the exponential
convergence of
the series. Those analytical
continuations in $q$
must be dealt with specifically. The physical example
of a field theory with  a chemical
potential falls clearly into this class.
%%%%%%%%%%%%%%%%%%%%%%%%%%%%%%%%
\subsection{Generalized Epstein zeta function in $d=2$}
%%%%%%%%%%%%%%%%%%%%%%%%%%%%%%%%
For completeness, let us write down the corresponding series when $p=2$
explicitly. They are, with $q>0$,
 \begin{eqnarray}
&& \hspace{-4mm}  \zeta_E(s;a,b,c;q) = -q^{-s}
+\frac{2\pi q^{1-s}}{(s-1) \sqrt{\Delta}}
 + \frac{4}{\Gamma (s)} \left[
\left( \frac{q}{a} \right)^{1/4}
 \left( \frac{\pi}{\sqrt{qa}} \right)^s \right. \nonumber \\ &&
\hspace*{-2mm} \times \sum_{n=1}^\infty
n^{s-1/2} K_{s-1/2} \left( 2\pi n \sqrt{\frac{q}{a}} \right)
 + \sqrt{\frac{q}{a}} \left(2\pi
\sqrt{\frac{a}{q\Delta}} \right)^s
 \sum_{n=1}^\infty n^{s-1} K_{s-1} \left( 4\pi n
\sqrt{\frac{a q}{\Delta}}\right)  \label{cse1}
 \\ && \hspace*{-10mm}+\left. \sqrt{\frac{2}{a}} (2\pi)^s
 \sum_{n=1}^\infty n^{s-1/2} \cos (\pi n b/a) \sum_{d|n} d^{1-2s} \,
 \left( \Delta + \frac{4aq}{d^2} \right)^{1/4-s/2}
K_{s-1/2}  \left( \frac{\pi n}{a} \sqrt{ \Delta + \frac{4aq}{d^2}}
\right) \right],\nonumber
\end{eqnarray}
where $\Delta = 4ac -b^2>0$, and, with $q=0$, the CS formula\cite{cs}
\begin{eqnarray}
&& \hspace*{-12mm} \zeta_E(s;a,b,c;0) = 2\zeta (2s)\, a^{-s} + \frac{2^{2s}
\sqrt{\pi}\, a^{s-1}}{\Gamma (s) \Delta^{s-1/2}} \,\Gamma (s
-1/2) \zeta (2s-1)
\nonumber \\ && \hspace*{-12mm}
+ \frac{2^{s+5/2} \pi^s }{\Gamma (s) \, \Delta^{s/2-1/4}\, \sqrt{a}}
\sum_{n=1}^\infty
 n^{s-1/2}
\sigma_{1-2s} (n) \,
 \cos (\pi n b/a) \,
K_{s-1/2}\left( \frac{\pi n}{a}
\sqrt{ \Delta} \right).
\label{cs1}
\end{eqnarray}
where
$
\sigma_s(n) \equiv \sum_{d|n} d^s,
$
 sum over the $s$-powers of the divisors of $n$.
 We observe that the rhs's of (\ref{cse1}) and
 (\ref{cs1}) exhibit a
simple pole at $s=1$, with common residue:
$
\mbox{Res}_{s=1}  \zeta_E(s;a,b,c;q) = \frac{2\pi}{\sqrt{\Delta}} =
\mbox{Res}_{s=1}  \zeta_E(s;a,b,c;0).
$
%%%%%%%%%%%%%%%%%%%%%%%%%%%%%%%%
\subsection{Truncated Epstein zeta function in $d=2$}
%%%%%%%%%%%%%%%%%%%%%%%%%%%%%%%%
The most involved case in the family of Epstein-like
 zeta functions corresponds to having to deal with
a {\it truncated} range. This
comes about when one imposes boundary conditions
of the usual Dirichlet or Neumann type.
 Jacobi's theta function identity and Poisson's summation
formula are then {\it useless} and no expression in terms of a
convergent series for the analytical continuation to
 values of $\Re s$ below the abscissa of convergence
can be obtained. The method must use then the  zeta function regularization
theorem\cite{eli11} and the best one gets
 is an {\it asymptotic} series. The issue of extending
 the CS formula, or the
most general expression we have obtained before, to this situation is not
an easy one (see, however, Ref.~[\refcite{cmpe1}]).
This problem has seldom (if ever)
been properly addressed in the literature.

As an example, let us consider the following series in
one dimension:
$
\zeta_G(s;a,c;q) \equiv \sum_{n=-\infty}^{\infty}
\left[ a(n+c)^2+q
\right]^{-s}, \ \Re s >1/2.
%\label{g1}
$
Associated with this zeta functions, but
 considerably more difficult to treat, is the
truncated series, with indices running from 0 to
$\infty$
\beq
\zeta_{G_t}(s;a,c;q) \equiv \sum_{n=0}^{\infty}
\left[ a(n+c)^2+q
\right]^{-s}, \quad  \Re s >1/2.
\label{g1a}
\eeq
 In this case the Jacobi identity is of no use.
 The  way to proceed  is  employing
specific techniques of analytic continuation
 of zeta functions.
There is no place to describe them here in detail.
The usual method involves three steps
\cite{eli11}. The first step is easy: to write the
initial series as a Mellin transform
$
 \sum_{n=0}^{\infty}
\left[ a(n+c)^2+q
\right]^{-s}= \frac{1}{\Gamma (s)} \sum_{n=0}^{\infty} \int_0^\infty
dt \, t^{s-1} \exp\left\{ -[ a(n+c)^2+q]t \right\}.
\label{g1b}
$
The second, to expand in power series part of the exponential,
leaving  a converging  factor:
$
 \sum_{n=0}^{\infty} \left[ a(n+c)^2+q \right]^{-s}=
\frac{1}{\Gamma (s)} \sum_{n=0}^{\infty} \int_0^\infty
dt \,  \sum_{m=0}^{\infty} \frac{(-a)^m}{m!} (n+c)^{2m} t^{s+m-1}
e^{-qt}. \label{g1c}
$
The third, and most difficult, step is to interchange the order of the
two summations ---with the aim to obtain a series of zeta functions---
what means transforming the second series into
an integral along a path on the complex plane, that has to be closed
into a circuit (the sum over poles inside reproduces the original series),
with a part of it being sent to
infinity. Usually, after interchanging the first
 series and the integral, there is a contribution of this part
of the circuit at infinity, what provides in the end an {\it additional}
contribution
to the trivial commutation (given by the zeta function regularization
theorem \cite{eli11}). More important, what one obtains in general
through this process is
{\it not} a convergent series of zeta functions, but an asymptotic
series (see e.g., the 4th and 6th references in [\refcite{zbooks1}]). That is, in our example,
$
\sum_{n=0}^{\infty} \left[ a(n+c)^2+q \right]^{-s}\sim
\sum_{m=0}^{\infty} \frac{(-a)^m\Gamma (m+s)}{m!\, \Gamma (s) \,
q^{m+s}} \zeta_H (-2m, c) + \, \mbox{additional terms}.
$
Being more precise, as outcome of the whole process
 we obtain the following result for the analytic
continuation of the zeta function\cite{elif1}
\begin{eqnarray}
&& \hspace*{-3mm} \zeta_{G_t}(s;a,c;q)
 \sim \left(\frac{1}{2} -c \right) q^{-s} + \frac{q^{-s}}{\Gamma (s)}
\sum_{m=1}^{\infty}
\frac{(-1)^m \Gamma (m+s)}{m!} \left( \frac{q}{a} \right)^{-m}
\zeta_H (-2m, c)  \label{if11} \\ &&   \hspace*{-8mm} +
\sqrt{\frac{\pi}{a}} \, \frac{\Gamma (s-1/2)}{2\Gamma (s)} q^{1/2 -s}
+\frac{2\pi^s}{\Gamma (s)} a^{-1/4-s/2} q^{1/4-s/2}
 \sum_{n=1}^\infty
n^{s-1/2} \cos (2\pi nc) K_{s-1/2} (2\pi n\sqrt{q/a}).  \nonumber
\end{eqnarray}
(Note that this expression reduces to Eq. (\ref{r01}) in the limit
$c \rightarrow 0$.)
The first series on the rhs is asymptotic\cite{eli11,8}.
Observe, on the other hand, the singularity structure of this zeta
function. Apart from the pole at $s=1/2$,  there is a whole sequence of
poles at the negative real axis, for $s= -1/2, -3/2,\ldots$,
with residua:
$
\mbox{Res}_{s=1/2-j} \zeta_{G_t}(s;a,c;q) = \frac{(2j-1)!!\, q^j}{
j!\, 2^j \sqrt{a}}, \ j=0,1,2, \ldots
$
The generalization of this to $p$ dimensions can be found in Ref.~[\refcite{cmpe1}].

\section{Direct physical applications}
%%%%%%%%%%%%%%%%%%%%%%%%%%%%%%%%
\subsection{Calculation of the Casimir energy density}
%%%%%%%%%%%%%%%%%%%%%%%%%%%%%%%

An application of the procedure is the calculation of the {\it Casimir energy
density} corresponding to a massless\cite{dowk89} or massive\cite{ke1} scalar field on a general, $d$ dimensional toroidal manifold.
In the spacetime ${\cal M} =$ {\bf R}
$\times \Sigma$, with $\Sigma = [0,1]^d/$$\sim$, which is
topologically equivalent
to the $d$ torus, the Casimir energy density for a massive scalar field
is given directly by Eq. (\ref{qpd1}) at $s=-1/2$, with $q=m^2$ (mass of
the
field), $\vec{b} =\vec{0}$, and $A$ being the matrix of the metric $g$ on
$\Sigma$, the general $d$-torus:
$
E^C_{{\cal M}, m} = \zeta_{g, \vec{0}, m^2} (s=-1/2).
$
The components of $g$ are, in fact, the
coefficients of
the different terms of the Laplacian, which is the relevant operator in
the Klein-Gordon field equation. The massless case is also obtained,
with the same specifications, from the corresponding formula  Eq.
(\ref{cspd}). In both cases no extra calculation needs to be done,
and the physical results follow from a mere
{\it identification} of the components of the matrix $A$ with those of the
metric tensor of the manifold in question\cite{ke1}. Very much related with
this application but more involved and ambitious  is the calculation of
vacuum  energy densities corresponding to spherical configurations and
the bag model (see
[\refcite{bekl1,wipf12}] and  references therein).
%%%%%%%%%%%%%%%%%%%%%%%%%%%%%%%%
\subsection{Effective action}
%%%%%%%%%%%%%%%%%%%%%%%%%%%%%%%
Another application consists in calculating the {\it determinant} of a
differential operator, say the Laplacian on a general $p$-dimensional
torus. A very important problem related with this issue is that
 of the multiplicative anomaly discussed before\cite{a1}.
 To this end the derivative of the zeta function at $s=0$ has to
be obtained. From Eq. (\ref{qpd1}), we get
\beq
 {\zeta'}_{A,\vec{c},q} (0) &=&
 \frac{4 (2q)^{p/4}}{\sqrt{\det A}}
\sum_{\vec{m} \in \mbox{\bf Z}^p_{1/2}}' \frac{\cos (2\pi
 \vec{m}\cdot \vec{c})}{ \left( \vec{m}^T A^{-1} \vec{m}
\right)^{p/4}} \, K_{p/2} \left( 2\pi \sqrt{2q \,
 \vec{m}^T A^{-1} \vec{m}}\right) \nn \\ && \hspace{5mm} + \left\{
\brr{ll} \dsp\frac{(2\pi )^{p/2} \Gamma (-p/2) q^{p/2}}{
\sqrt{\det A}}, & p \ \mbox{odd}, \\
\dsp\frac{(-1)^k(2\pi )^k q^k}{k!\,
\sqrt{\det A}} \, \left[ \Psi (k+1) +\gamma -\ln q \right], &
p=2k \ \mbox{even}, \err \right.
%\nn % \label{qpdd1}
\eeq
and, from here, det $A$ = exp $-\zeta_A'(0)$. For $p=2$, we have
explicitly:
\beq
&& \hspace*{-5mm} \det A(a,b,c;q) = e^{2\pi (q- \ln q)/\sqrt{\Delta}}
 \left( 1-
e^{-2\pi \sqrt{q/a}} \right) \exp \left\{ -4 \sum_{n=1}^\infty
\frac{1}{n} \left[ \sqrt{\frac{a}{q}} \ K_1 \left( 4\pi n \sqrt{
\frac{aq}{\Delta}} \right) \right. \right. \nn \\ && \hspace*{10mm}+
\left. \left. \cos (\pi n b /a) \sum_{d|n} d \, \exp \left(-
\frac{\pi n}{a} \sqrt{ \Delta + \frac{4aq}{d^2}} \right) \right]
\right\}. \eeq
In the homogeneous case (CS formula) we obtain for the determinant:
\beq
\hspace*{-3mm}\det A(a,b,c) = \frac{1}{a} \exp \left[ -4 \zeta'(0) -
\frac{\pi \sqrt{\Delta}}{6a} -4 \sum_{n=1}^\infty \frac{\sigma_1(n)}{n}
\cos (\pi n b /a) e^{-\pi n \sqrt{\Delta} /a} \right],
\eeq
or, in terms of the Teichm\"uller coefficients, $\tau_1$ and $\tau_2$,
of the metric tensor (for
the metric, $A$, corresponding to the general torus in two dimensions):
\beq
\hspace*{-3mm}
\det A(\tau_1, \tau_2) = \frac{\tau_2}{4\pi^2|\tau|^2} \exp \left[ -4
\zeta'(0) -
\frac{\pi \tau_2}{3|\tau|^2} -4 \sum_{n=1}^\infty \frac{\sigma_1(n)}{n}
\cos \left(\frac{2\pi n \tau_1}{|\tau|^2} \right) e^{-\pi n
\tau_2/|\tau|^2} \right]. \nn \\
 \eeq

%%%%%%%%%%%%%%%%%%%%%%%

\section{Future Perspectives: Operator Regularization}
The Operator Regularization (OR) approach, due originally to
D. G. C. McKeon and T. N. Sherry\cite{ks1} is considered as a genuine generalization of the zeta regularization approach. Its main aim is to extend zeta regularization, so effective at one-loop order\cite{eno1}, to higher loops. It has a distinct advantage over other competing procedures, in that it can be used with formally
non-renormalizable theories, as shown in [\refcite{mtks1,shie1}].
A further feature of this approach is that divergences are not reabsorbed, each one is removed and replaced by an arbitrary factor. Indeed, operator regularization (OR)
does not cure the non-predictability problem of non-renormalizability,
but an advantage of the method is that the initial Lagrangian does not need to be extended with
the addition of extra terms.
The OR scheme is governed by the identity:
\beq H^{-m} = \lim_{\epsilon \to 0} \frac{d^n}{d\epsilon^n}
\left[1 + \left( 1 + \alpha_1\epsilon + \alpha_2\epsilon^2+ \ldots
+ \alpha_n\epsilon^n\right)\frac{\epsilon^n}{n!}\, H^{-\epsilon-m}\right],
\eeq
where the $\alpha_i$'s are arbitrary, and it is enough that the degree of regularization
 is equal to the loop order, $  n$.

Two separate aspects of the procedure are, first the regularization itself and, second,
the  analytical continuation, where divergences are replaced by arbitrary factors. Thus, the
effect of OR is in the end replace the
divergent poles by arbitrary constants, as
\beq \frac{1}{\epsilon^n} \longrightarrow \alpha_n, \eeq
to yield the finite expression
\beq H^{-m} = \alpha_n c_{-n} + \cdots + \alpha_1 c_{-1} + c_0. \eeq

\subsubsection{Generalization and further extensions}
The OR method can be generalized to {\it multiple operators}, as in multi-loop cases
\beq  H^{-m_1} \cdots H^{-m_r}  &=& \lim_{\epsilon \to 0} \frac{d^n}{d\epsilon^n}
\left[1 + \left( 1 + \alpha_1 \epsilon + \alpha_2\epsilon^2+ \cdots
+ \alpha_n\epsilon^n\right)\right. \nn \\ &&
\times \, \frac{\epsilon^n}{n!} \, \left. H^{-\epsilon-m_1}  \cdots H^{-\epsilon-m_r}\right]
\eeq
Further extensions of the procedure have been proposed. Let us recall that
OR was first introduced in the context of the
Schwinger approach,
\beq \ln H = -\lim_{\epsilon \to 0} \frac{d^n}{d\epsilon^n}
\left(\frac{\epsilon^{n-1}}{n!}\, H^{-\epsilon}\right), \eeq
which is known to be equivalent to the  Feynman one
\beq H^{-m} = \lim_{\epsilon \to 0} \frac{d^n}{d\epsilon^n}
\left(\frac{\epsilon^{n}}{n!}\, H^{-\epsilon-m}\right).\eeq
The Schwinger form can be transformed into the Feynman one, as
\beq H^{-m} = \frac{(-1)^{m-1}}{(m-1)!}\, \frac{d^m}{d H^m}\, \ln H \eeq
Equivalence with dimensional regularization can be established in many cases, but
not always. Problems, the main one being unitarity, may appear (see [\refcite{reb1}]).
To start with, its naive application to obtain finite amplitudes breaks unitarity.

A definite advantage of the procedure is that, actually, no symmetry-breaking regulating parameter is ever inserted into the initial Lagrangian\cite{clmks1}. One can use Bogoliubov's recursion formula in order to show how to construct a consistent OR operator, and unitarity is upheld by employing
a generalized evaluator consistently including lower-order quantum corrections to the quantities of interest. Unitarity requirements lead to {\it unique} expressions for quantum field theoretic quantities, order by order in $\hbar$. This fact has been proven in many cases (as for the $\Phi^4$ theory at two-loop order, etc.). But I should say that, to my knowledge, a universal proof of this issue is actually still missing.

A final comment is in order. Using a BPHZ-like scheme, as the above one turns out to be, in the end, essentially reintroduces counterterms into the procedure, since they are actually hidden in the subtractions taking place at each step. In this way, the simplicity of the original zeta function regularization procedure, as described in the previous sections, and which is one of its main characteristics, is absent in the extended, operator regularization method.

\section*{Acknowledgments}
The author is indebted with his several collaborators in this research during a number of years, in special with A.A. Bytsenko, G. Cognola, J. Haro, K. Kirsten, S. D. Odintsov, A. Romeo, A. Saharian, and S. Zerbini. The author acknowledges financial support by the European Science Foundation (ESF) within the activity ``New Trends and Applications of the Casimir Effect" (www.casimir-network.com).
This investigation has been also supported in part by MICINN (Spain), project
and FIS2010-15640, by the CPAN Consolider Ingenio Project, and by AGAUR (Generalitat
de Ca\-ta\-lu\-nya), contract 2009SGR-994. Research was partly carried out while
the author was on leave at the Department of Physics and Astronomy,
Dartmouth College, 6127 Wilder Laboratory, Hanover, NH 03755, USA.

\end{document}